# Distinctive momentum dependence of the band reconstruction in the nematic state of FeSe thin film

Y. Zhang[1,2], M. Yi[1,3], Z.-K. Liu[1,3], W. Li[1], J. J. Lee[1,3], R. G. Moore[1], M. Hashimoto[4], N. Masamichi[5,6], H. Eisaki[5,6], S.-K. Mo[2], Z. Hussain[2], T. P. Devereaux[1,3], Z.-X. Shen[1,3]* & D. H. Lu[4]*

[1]*Stanford Institute for Materials and Energy Sciences, SLAC National Accelerator Laboratory, 2575 Sand Hill Road, Menlo Park, California 94025, USA*

[2]*Advanced Light Source, Lawrence Berkeley National Laboratory, Berkeley, California 94720, USA*

[3]*Geballe Laboratory for Advanced Materials, Departments of Physics and Applied Physics, Stanford University, Stanford, California 94305, USA*

[4]*Stanford Synchrotron Radiation Lightsource, SLAC National Accelerator Laboratory, 2575 Sand Hill Road, Menlo Park, California 94025, USA*

[5]*National Institute of Advanced Industrial Science and Technology, Tsukuba, Ibaraki 305-8568, Japan*

[6]*JST, Transformative Research-Project on Iron Pnictides, Tokyo, 102-0075, Japan*

\* *To whom correspondence should be addressed:* zxshen@stanford.edu *and* dhlu@slac.stanford.edu

Nematic state, where the system is translationally invariant but breaks the rotational symmetry, has drawn great attentions recently due to experimental observations of such a state in both cuprates and iron-based superconductors. The mechanism of nematicity that is likely tied to the pairing mechanism of high-Tc, however, still remains controversial. Here, we studied the electronic structure of multilayer FeSe film by angle-resolved photoemission spectroscopy (ARPES). We found that the FeSe film enters the nematic state around 125 K, while the electronic signature of long range magnetic order has not been observed down to 20K indicating the non-magnetic origin of the nematicity. The band reconstruction in the nematic state is characterized by the splitting of the $d_{xz}$ and $d_{yz}$ bands. More intriguingly, such energy splitting is strong momentum dependent with the largest band splitting of ~80meV at the zone corner. The simple on-site ferro-orbital ordering is insufficient to reproduce the nontrivial momentum dependence of the band reconstruction. Instead, our results suggest that the nearest-neighbor hopping of $d_{xz}$ and $d_{yz}$ is highly anisotropic in the nematic state, the origin of which holds the key in understanding the nematicity in iron-based superconductors.

High-Tc superconductivity always occurs in the proximity to the symmetry breaking states, whose origin is intimately related to the pairing mechanism of superconductivity. Among all these symmetry breaking states, the nematic state, in which electrons self-organize unidirectionally and break the rotational symmetry without breaking the translational symmetry, has recently drawn great attentions[1,2]. It was found in the proximity to high-$T_C$ superconductivity in both cuprates and iron-based superconductors[3-9], and the quantum fluctuation near the nematic quantum critical point was proposed to be critical for the high-$T_C$[10,11].

In iron-pnictide superconductors, the nematic state develops simultaneously with a structural transition from tetragonal to orthorhombic, where the lattice breaks the C4 rotational symmetry[2]. A magnetic transition into a collinear antiferromagnetic (CAF) order follows at a lower temperature or simultaneously with the structural transition, further breaking the translational symmetry[12]. In order to explain the nematic state, most theories emphasize the importance of orbital/spin fluctuations or their strong coupling, in connection with the multi-orbital and correlated nature of iron-based superconductors[13-19]. However, the key debates among these theories lie in which fluctuation dominates at high temperature and what relationship the nematic and magnetic states have. The answer to such debates is crucial for understanding the nature of the competing phases and needs to be settled experimentally. However, clear experimental delineation of the nematic state has always been challenging. In most cases, the nematic and magnetic states strongly intertwine with each other. The freezing of spin fluctuations and CAF domains were found even above the magnetic transition temperature $(T_N)$[20], preventing us from probing the intrinsic properties of the nematic state.

In this paper, we present angle-resolved photoemission spectroscopy (ARPES) studies on the 35 monolayers (35ML) FeSe film grown on $SrTiO_3$. Our data show that the 35ML FeSe film is a cleanly delineated system with a nematic transition at ~125 K as characterized by the onset of strong anisotropy between $d_{xz}$ and $d_{yz}$ bands, while the signature of long-range magnetic order, i.e., band folding, has not been observed even at the lowest measurement temperature (20 K). More intriguingly, we found that the energy splitting between $d_{xz}$ and $d_{yz}$ bands in the nematic state shows distinctive distribution in the momentum space. Specifically, from the zone center (Γ) to the zone corner (M), the splitting energy of the electronic bands first decreases, then increases, and finally achieves the maximal value of ~80 meV at M. Our results thus highlight the dominating role of the bands around the zone corner in driving the nematic state, which puts strong constrains on the theories. The huge splitting energy between $d_{xz}$ and $d_{yz}$ bands and the absence of the magnetic order favor the orbital-ordering scenario. However, the nontrivial momentum dependence of the band reconstruction excludes the simple on-site ferro-orbital ordering as the driving force of the nematicity. The huge anisotropy of the nearest-neighbor hopping of the $d_{xz}$ and $d_{yz}$ orbitals should play a more important role.

# Result

## Absence of magnetic ordering in FeSe thin film

Most undoped iron-pnictide superconductors exhibit a CAF order, which breaks the translational symmetry of the system[12]. As a result, the unit cell rotates 45 degrees and its size doubles. The electronic manifestation for such a translational symmetry breaking is the band folding in the momentum space with respect to the magnetic zone boundary. We take NaFeAs as an example, whose structural transition temperature ($T_S$) and $T_N$ are around 56 and 46 K, respectively[22]. As shown in Fig. 1a, the Fermi surface of NaFeAs at 70 K consists of hole pockets around Γ and electron pockets around M[23, 24]. When the temperature is below $T_N$, the bands around Γ and M fold onto each other (Fig. 1b), and the Fermi surface sheets reconstruct so that they are symmetric with respect to the magnetic Brillouin zone (BZ) boundary (Fig. 1a). The small discrepancy of the Fermi surface between Γ and M is due to the different photoemission intensity between the main bands and the folded bands in the photoemission process as well as the photoemission matric element effects. We note that, without the translational symmetry breaking, the size of the unit cell remains the same and the bands do not fold in the nematic state. Therefore, the band folding observed here could be only caused by the translational symmetry breaking, which is the spectroscopic evidence for long-range magnetic order in NaFeAs.

In contrast, such a band folding behavior is clearly absent in the 35ML FeSe film. Figure 1c shows the Fermi surface mappings taken on 35ML FeSe film. The Fermi surface at 160 K consists of small hole and electron pockets, which is similar to that of NaFeAs. However, at 20 K, the Fermi surface of 35ML FeSe only reconstructs around the M point and forms four intense propeller-like pockets around M, which appears to be very different from the Fermi surface around Γ. Such a strong discrepancy indicates the absence of band folding, which is further confirmed by measured dispersion taken along the Γ-M direction (Fig. 1d). Unlike the high temperature spectra where only one band dominates, the low temperature spectra show multi-band behavior, which is more evident around M. However, none of the bands observed at 20 K could be attributed to the band folding. Besides the absence of band folding, further evidences come from the absence of the spin density wave (SDW) gap. In contrast to NaFeAs, where the band dispersions break into sections in the SDW state due to the SDW gap opening (Fig. 1b), the band disperse continuously and no SDW gap can be detected in FeSe (Fig. 1d). All these results unambiguously demonstrate the absence of magnetic order in 35ML FeSe film, which is also consistent with the Nuclear Magnetic Resonance (NMR) and Mössbauer measurements on bulk FeSe showing no static magnetic order at the lowest temperature[25-27].

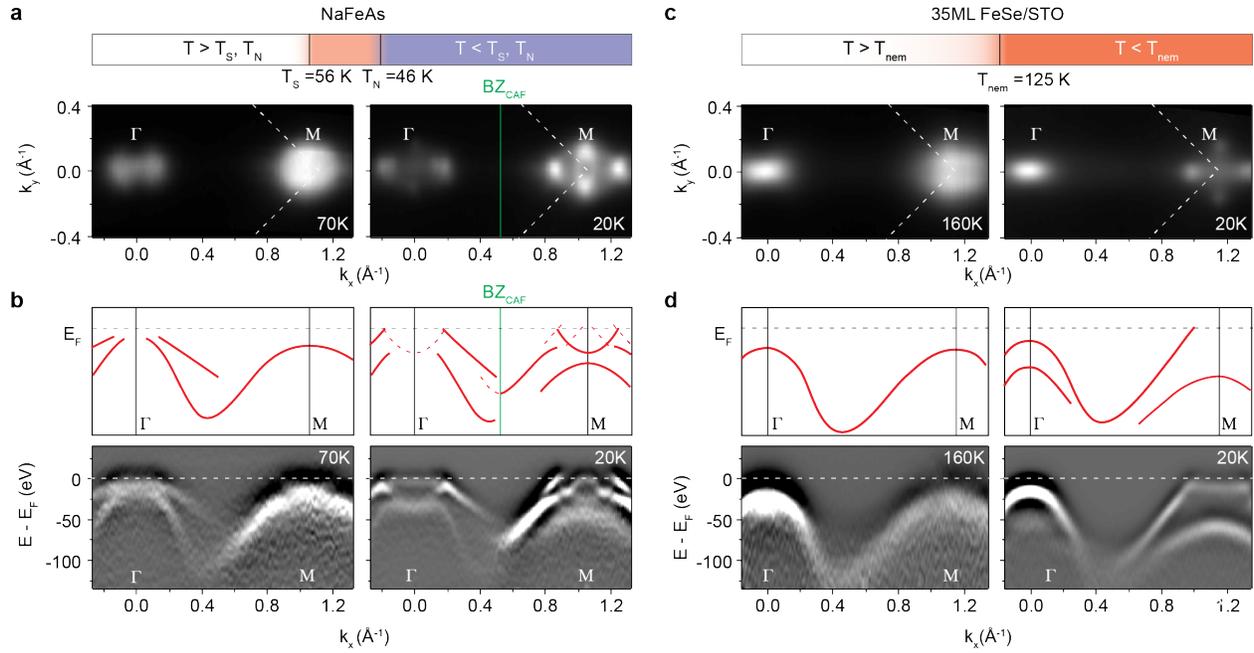

**Fig. 1: Absence of magnetic ordering in FeSe thin film.** (a) The Fermi surface mapping taken at 70 and 20 K in NaFeAs. The Brillouin zone (BZ) boundary in the magnetic state is shown by green sold lines. The structural transition temperature and magnetic transition temperature are abbreviated as $T_S$ and $T_N$, respectively. (b) The second derivative photoemission intensity distribution taken along the Γ-M direction. The main (solid line) and folded (dashed line) bands are illustrated in the upper panels. (c) and (d) are the corresponding data taken in 35ML FeSe film at 160 and 20 K, respectively. The nematic transition temperature is abbreviated as $T_{nem}$.

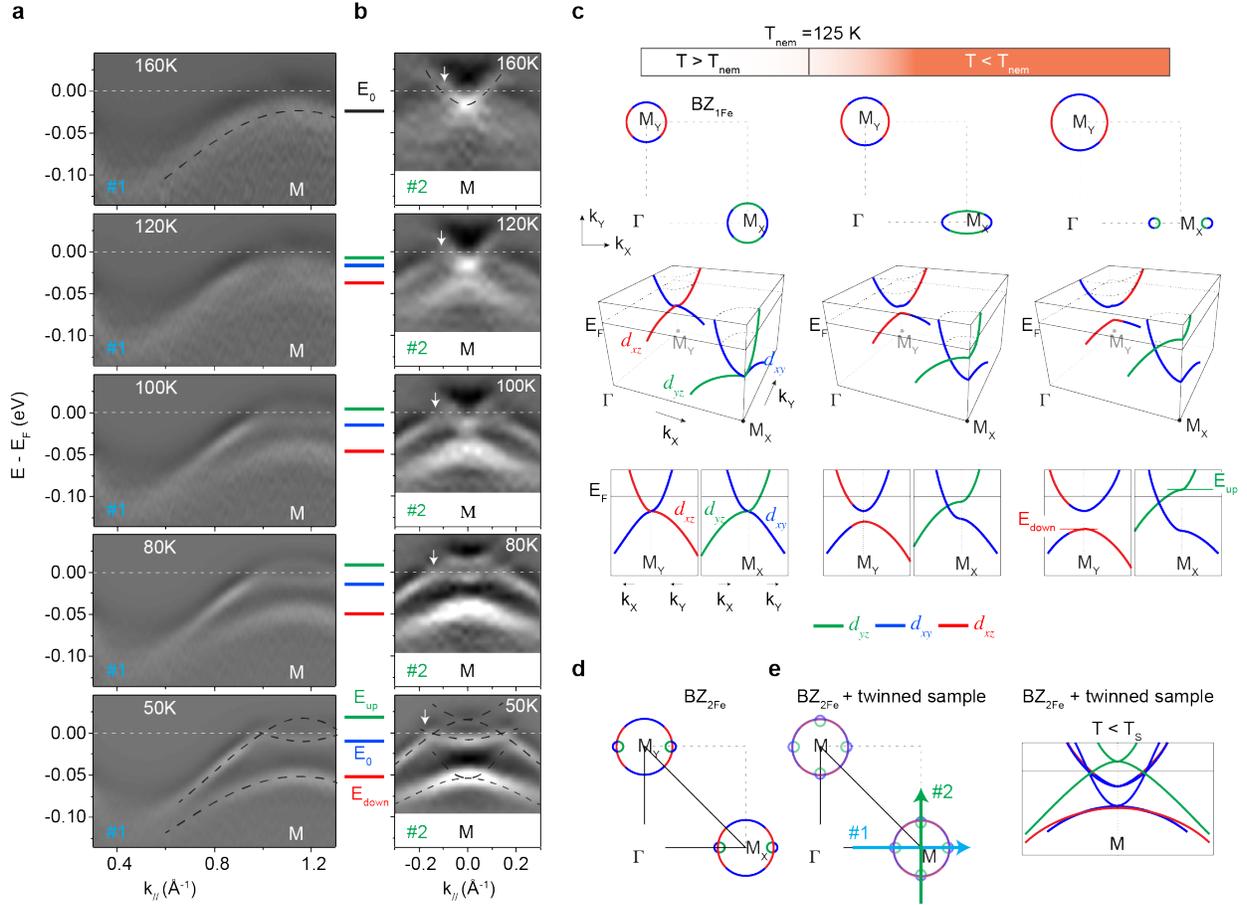

**Fig. 2: Band reconstruction in the nematic state.** (a) Temperature dependence of the second derivative photoemission intensity distribution taken from 35ML FeSe film along the Γ-M direction (cut #1 as indicated in panel e). (b) is the same as panel a, but taken along the perpendicular cut #2 direction as indicated in panel e. The solid lines with different colors mark the energy positions of either the band tops of hole-like bands or bottoms of electron-like bands. (c) Illustration of the reconstruction of the Fermi surface (upper panel) and the band structure (middle and lower panels) around the M point in one-Fe BZ of 35ML FeSe film. For simplicity, we neglect the hole bands and associated hole pockets near the zone center the Γ point. (d) Illustration of the low temperature Fermi surface in two-Fe BZ by folding the corresponding bands between $M_X$ and $M_Y$. (e) Illustration of the Fermi surface and the band structure around M

in twinned sample by overlapping the bands in two perpendicular domains. The results are consistent with the observed band dispersions.

**Band reconstruction in the nematic state**

The multilayer FeSe film resembles the bulk FeSe and shows C4 symmetry breaking at low temperature[28]. Without the interference of the folded bands and SDW gap associated with magnetic order, the band reconstruction observed in 35ML FeSe film is simpler and reflects the intrinsic properties of the nematic state. Figures 2a and 2b show the temperature evolution of the photoemission spectra along the Γ-M and its orthogonal direction in 35ML FeSe film. The band reconstruction at the M point is most pronounced and could be characterized by tracking both the band tops of the hole-like bands near M in cut #1 (Fig. 2a) and the band bottoms of the electron-like bands in cut #2 (Fig. 2b). The band tops and bottoms are degenerate at $E_0$ at 160 K, respecting the C4 symmetry of the tetragonal lattice. Upon lowering the temperature, the hole and electron bands shift consistently and splits into three separate branches: one branch ($E_{up}$) shift upwards above the Fermi energy ($E_F$) and another branch ($E_{down}$) shift downwards to higher binding energy. In the middle, a shallow electron band was observed at low temperature, whose band bottom ($E_0$) is nearly unchanged with temperature.

The observed band reconstruction could be explained by the orbital-dependent band shift and sample twinning effect. As shown in Fig. 2c, we first consider the band structure in one-Fe BZ. For simplicity, we only focus on the band structure near the M point where the band reconstruction is most apparent. The $d_{yz}$ and $d_{xz}$ bands hybridize with the $d_{xy}$ bands near $M_X$ and $M_Y$, respectively, forming two pairs of electron and hole-like bands[29]. In the nematic state, the $d_{yz}$ band shifts up around the $M_X$ point. As a result, the electron pocket shrinks along one direction and eventually turns into two small Fermi pockets. On the contrary, around the $M_Y$ point, the $d_{xz}$ band shifts downwards and opens a hybridization gap with the $d_{xy}$ band, which enlarges the electron pocket at the $M_Y$ point. We note that, except for the $d_{yz}$ electron band, the Fermi crossings ($k_F$'s) of all other electron bands expand as shown by the white arrows in Fig. 2b. This observation indicates that, similar to the $d_{xz}$ electron band, the $d_{xy}$ band also participates in the band reconstruction in the nematic state with its $k_F$ enlarges at low temperature. After folding these bands, we then obtained the reconstructed Fermi surface in the two-Fe BZ (Fig. 2d). Moreover, due to the sample twinning effect, the observed band structure originates from two perpendicular domains[4]. Considering this effect, Fig. 2e reproduces the observed propeller-like Fermi pockets and the band splitting at the M point very well. We note that part of the large circular electron pocket in Fig. 2e is constructed by the $d_{xy}$ orbital, whose photoemission matrix elements is much weaker than the $d_{xz}$ and $d_{yz}$ orbitals[5]. Therefore, the circular electron pocket is missing in the Fermi surface mapping in Fig. 1c.

The splitting behavior of the $d_{xz}$ and $d_{yz}$ bands is consistent with previous ARPES studies on the detwinned Ba(Fe$_{1-x}$Co$_x$)As$_2$ and NaFeAs[5, 23, 24]. However, we should emphasize that, due to the absence of band folding that is associated with the magnetic order, the band reconstruction is much simpler and more distinguishable in FeSe. For example, the whole $d_{yz}$ band was observed shifting up above $E_F$ in FeSe film, while in FeAs compounds, only a small section of this band can be observed due to the band folding and gap opening. Moreover, our data on the thin film are also consistent with the ARPES studies on the electronic structure of bulk FeSe[30, 31]. While the consistency indicates the bulk-like behavior of the 35ML FeSe film, the high quality of the film grown by the MBE method enables us to resolve the detailed band reconstruction in the nematic state for the first time.

**Nontrivial momentum dependence of the band reconstruction**

FeSe film is an ideal system for us to quantitatively study the band reconstruction in the nematic state. One intriguing behavior in Fig. 2a is the strong momentum dependence of the band reconstruction. The energy separation between the $d_{xz}$ and $d_{yz}$ bands decreases significantly when moving away from the M point. To see it more clearly, we overlaid the high temperature band dispersion that is extracted from Fig. 3a (the red dashed lines) on top of the low temperature spectra image, as shown in Fig. 3b. The band reconstruction at M is much more pronounced than that at Γ. Figure 3c shows the detailed temperature evolution of the energy distribution curves (EDCs) taken at five different momenta as indicated in Fig. 3b. The band positions at each temperature were determined through the peak fitting and plotted in Fig. 3d. All the bands start to shift at ~125 K, which we interpret as the nematc transition temperature ($T_{nem}$) for 35ML FeSe film as this is the temperature symmetry breaking between the $d_{xz}$ and $d_{yz}$ orbitals. This temperature is higher than the $T_S$ (90 K) in bulk FeSe, which could be due to the lattice strain of the film[32]. In the nematic state, the magnitude of the band shift is strongly momentum-dependent and changes non-monotonically from Γ to M (Fig. 3d). More specifically, it is around 10 meV at Γ (#1), and decreases to its minimal value of ~4 meV at the momentum slightly away from Γ (#2), then increases towards M (#3 and #4), and finally reaches its maximal value of 30~45meV at M (#5).

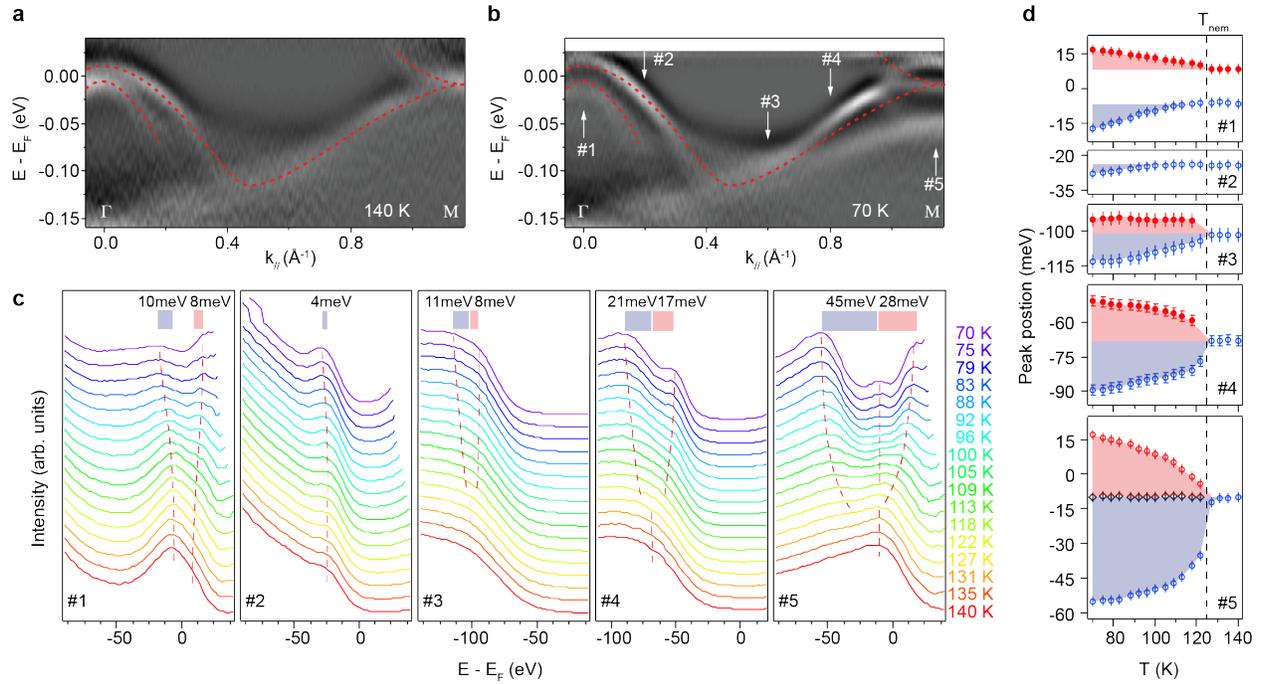

**Fig. 3: Nontrivial momentum dependence of the band reconstruction.** (a) The second derivative photoemission intensity distribution taken in 35ML FeSe film along the Γ-M direction at 140 K. (b) is the same as panel a, but taken at 70 K. The red dashed lines show the high temperature band dispersion extracted from panel a. (c) The temperature dependence of the energy distribution curves (EDCs) taken at five different momenta, after the division of Fermi-Dirac function. The peak positions are determined via a combination of spectral weight maximum and second derivative curve minimum. The top red and blue bars illustrate the energy

scale of the band shift. (d) The temperature dependence of the band positions extracted from panel c.

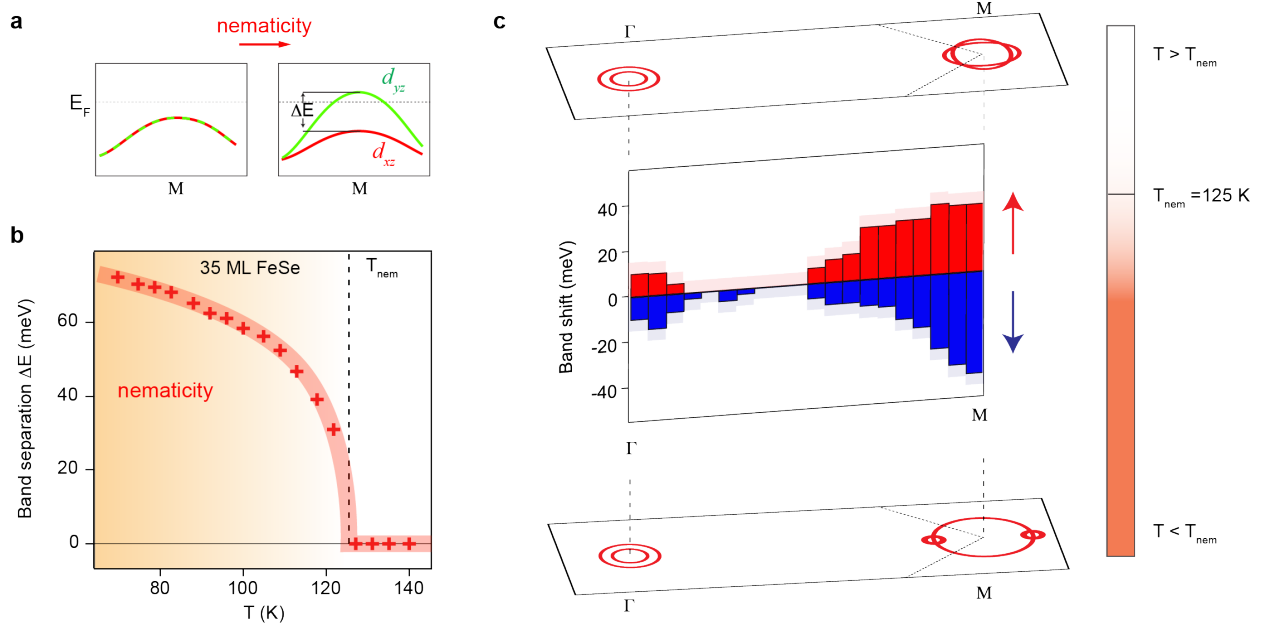

**Fig. 4: Summary of the band reconstruction in the nematic state.** (a) Energy splitting of $d_{xz}$ and $d_{yz}$ bands in the nematic state. (b) Temperature dependence of the band separation in 35ML FeSe. (c) The summary of the Fermi surface reconstruction and the momentum dependence of the band shift along Γ-M direction through in the nematic state.

**Discussion**

The detailed momentum dependence of the band shift in the nematic state is summarized in Fig. 4. The observed band reconstruction and its distinctive distribution in the momentum space put strong constrains on theories for understanding the nature of nematicity. Under the spin-nematic scenario[18, 19], both the nematic and magnetic states share the same origin, namely the spin fluctuation. It has been proposed that the spin fluctuation could break the C4 rotational symmetry by peaking at (0, π) rather than (π, 0), resulting in a spin-nematic state. Such a state is located slightly above the magnetic ordered state and is responsible for the structural transition through magnetic-elastic coupling. Our results do not favor such a spin-nematic scenario, because no signature of long-range magnetic ordering, i.e., band folding, was observed in FeSe down to the lowest measurement temperature (20K), even though the $T_{nem}$ is as high as 125 K (Fig. 3c) and the band splitting energy around M is as large as 80 meV (Fig. 4). One may argue that the FeSe system might have strong spin fluctuation without long-range magnetic order, such as the spin liquid state. However, the existence of spin liquid requires fine balance of the parameters, for example the perfect two-dimensionality or the lack of Fermi surface nesting.  These are not applicable for FeSe, because its electronic structure shows moderate three-dimensionality and the Fermi surface at high temperature is very similar to that of iron-arsenides exhibiting good nesting between Γ and M (Fig. 1c). Furthermore, the spin liquid state is unstable and will form magnetic order if any of the symmetry breaks, while in FeSe, the system breaks the C4 rotational symmetry in a dramatic way leaving no room for the existence of spin liquid. Based on all these facts, the electronic nematicity here cannot be attributed to the spin-nematicity. This is further supported by recent NMR and thermodynamic study on FeSe single crystal, which also points out the non-magnetic origin of the nematic state[26, 27].

Ruling out the spin-nematicity scenario, let us consider the possibility of orbital ordering. In orbital-ordering scenarios, the nematic and magnetic states are considered separately[13-15]. The orbital fluctuation dominates at high temperature and breaks the C4 rotational symmetry by triggering a $d_{xz}/d_{yz}$ ferro-orbital ordering. Then, at a lower temperature, the orbital ordering and C4 symmetry breaking further enhance the spin fluctuation and its anisotropy, which results in the magnetic transition with sufficiently strong coupling between spin and orbital/lattice degree of freedom. Such a scenario could naturally explain the absence of magnetic ordering in FeSe assuming a weak coupling between spin and orbital/lattice. On the other hand, the existence of a simple ferro-orbital ordering would lead to an energy difference between $d_{xz}$ and $d_{yz}$ orbitals and such energy difference should be momentum independent if only the on-site interactions are considered. However, as shown in Fig. 4, the band shift at $\Gamma$, where the orbital character of bands is pure and atomic-like, gives the maximal on-site energy difference between the $d_{xz}$ and $d_{yz}$ orbitals of ~20 meV, which is much smaller than the splitting energy of ~80 meV at M. Therefore, the on-site ferro-orbital ordering itself could not fully account for the nontrivial momentum dependence of the band reconstruction observed here. Moreover, we noticed that the $d_{xy}$ bands also exhibit noticeable reconstruction in the nematic state, which is usually ignored in two-band theoretical models[13-15]. From these perspectives, our results thus suggest that the nematicity in multilayer FeSe is not triggered purely by the simple on-site ferro-orbital ordering, while a more complex and itinerant form of orbital ordering that involves all three orbitals may need to be considered.

The most notable reconstruction of electronic structure in the nematic state is the energy splitting of the $d_{xz}$ and $d_{yz}$ bands and its non-trivial momentum dependence. According to theoretical calculations, such energy splitting could be due to the strong anisotropy in the nearest-neighbor intra-orbital hopping of the $d_{xz}$ and $d_{yz}$ orbitals[15, 16], which leads to a momentum-dependent band splitting with its maximum at the zone corner and minimum at the zone center[16, 17]. The energy scale of the anisotropic hopping could be then determined by the energy splitting at the zone corner[16], which is around 80 meV. Such an energy scale is much larger than the 20 meV energy scale of the ferro-orbital ordering observed at the zone center, which suggest that the anisotropic hopping of $d_{xz}$ and $d_{yz}$ orbitals plays a more important role in driving the nematicity. The origin of such strong anisotropy of hopping requires further understanding. It cannot originate purely from the lattice distortion, because the change of lattice constant is too small to account for such a large energy scale[5]. The coupling among the spin fluctuation, orbital ordering, and instability of band structure should be taken into account.

**Methods**

FeSe films were grown on the Nb doped $SrTiO_3$ substrate by the molecular beam epitaxy (MBE) method with a thickness of 35 ML, as described in Ref. 21. ARPES measurements were performed at the beamline 5-4 of Stanford Synchrotron Radiation Lightsource (SSRL) and the beamline 10.0.1 of Advanced Light Source (ALS). All data were taken with Scienta R4000 electron analyzers. The overall energy resolution was 5~10 meV depending on the photon energy, and the angular resolution was 0.3°. For the ARPES measurements at SSRL, the films were transferred from the growth chamber to ARPES chamber via a vacuum suitcase with the pressure better than $1\times10^{-9}$ torr. For the ARPES measurements at ALS, each film was capped with a Se layer of 25 nm thickness to protect the thin film during sample transfer. The film was then heated up to 400 °C to decap the Se capping layer in the ARPES chamber. All the samples were measured in ultrahigh vacuum with a base pressure better than $3\times10^{-11}$ torr.

**Acknowledgements**

ARPES experiments were performed at the Stanford Synchrotron Radiation Lightsource and the Advanced Light Source, which are both operated by the Office of Basic Energy Sciences, U.S. Department of Energy. The Stanford work is supported by the US DOE, Office of Basic Energy Science, Division of Materials Science and Engineering, under award # DE-AC02-76SF00515.

**Author Contributions**

Y. Z. and D. H. L. conceived the project. Y. Z., M. Y., Z. –K. L. and W. L. took the measurements. Y. Z. analyzed the data. W. L., J. J. L. and R. G. M. grew the films. N. M. and H.

**Author Information**

The authors declare no competing financial interests. Correspondence and request for materials should be addressed to D. H. Lu (dhlu@slac.stanford.edu) and Z.-X. Shen (zxshen@stanford.edu).